\documentclass[twocolumn,pre,floatfix]{revtex4}
\usepackage{amsmath,amssymb,graphicx}

\newcommand{\eq}[1]{Eq.(\ref{#1})}
\newcommand{\fig}[1]{Fig.~\ref{#1}}

\newcommand\ac{Acta. Crystallogr.}
\newcommand\aicej{Amer. Inst. Chem. Eng. J.}
\newcommand\acp{Adv. Chem. Phys.}

\newcommand\dt{Dalton Trans.}
\newcommand\el{Europhys. Lett.}
\newcommand\fd{Faraday Discuss.}
\newcommand\jmr{J. Mater. Res.}
\newcommand\lang{Langmuir}
\newcommand\nl{Nano Lett.}
\newcommand\nps{Nature Phys. Sci.}
\newcommand\jpca{J. Phys. Chem. A}
\newcommand\jpcb{J. Phys. Chem. B}
\newcommand\jpcm{J. Phys. Cond. Mat.}

\newcommand\pnasu{Proc. Natl. Acad. Sci. USA}
\newcommand\ptrsla{Philos. Trans. Roy. Soc. Lond. A}
\newcommand\rpp{Rep. Prog. Phys.}
\newcommand\sci{Science}
\newcommand\vir{Virology}

\begin{document}

\title{Structural trends in clusters of quadrupolar spheres}

\author{Mark A.~Miller}
\author{James J.~Shepherd}
\author{David J.~Wales}
\affiliation{University Chemical Laboratories, Lensfield Road,
Cambridge CB2 1EW, United Kingdom}

\date{\today}

\begin{abstract}
The influence of quadrupolar interactions on the structure of small clusters is
investigated by adding a point quadrupole of variable strength to the Lennard-Jones
potential.  Competition arises between sheet-like arrangements of the particles,
favoured by the quadrupoles, and compact structures, favoured by the isotropic
Lennard-Jones attraction.  Putative global potential energy minima 
are obtained for clusters of up to 25 particles using the basin-hopping algorithm.
A number of structural motifs and growth sequences emerge, including star-like
structures, tubes, shells and sheets.  The results are discussed in the context of
colloidal self-assembly.
\end{abstract}

\maketitle

\section{Introduction}

The structure adopted by a collection of particles is ultimately governed by
the energetic interactions between the particles.  It is therefore natural to
ask what sort of structures are favoured by a given set of interactions, and also whether
interactions can be chosen or manipulated in order to produce a particular target
structure.  The scope of both these questions is growing increasingly broad as it
becomes possible to exert ever greater control over the shape and form of the interactions
between molecular and colloidal building blocks \cite{Glotzer04b}. The motivation
for seeking deeper understanding in these areas is the desire to design novel
materials and supramolecular structures with unusual and useful properties.
\par
Considerable structural variety is possible even for spherical particles and isotropic
interparticle potentials.  Simple van der Waals interactions between
inert gas atoms promote near-spherical, highly-coordinated structures, favouring
icosahedral packing for clusters \cite{Hoare71a,Echt81a} and close-packed crystals
in the bulk.  However, these structures can be suppressed by introducing a repulsive
barrier into the pair potential at a distance close to $\sqrt{2}$ times the nearest
neighbour separation \cite{Dzugutov92a}.  In clusters, potentials of this form further promote
icosahedral local order, but lead to polytetrahedral structures \cite{Doye01b},
or to less compact shapes that are either elongated
or contain holes \cite{Doye01a}.  Local maxima in the pair potential can arise, 
for example, from the combination
of short-range depletion attraction between colloidal particles with partially screened long-range
Coulomb repulsion \cite{Mossa04a}, in which case the accumulated charge of a growing cluster
can effectively limit the size of aggregate that forms.  It is also possible to favour less
highly-coordinated order by careful design of isotropic pair potentials.  For example,
inverse statistical mechanical techniques can be used to derive isotropic potentials that
favour crystalline but non-close-packed bulk structures, such as the diamond and wurtzite
lattices \cite{Rechtsman07b}.
\par
A vast range of structures become possible when either non-spherical
particles or anisotropic interactions are considered.  Evolution has selected molecules
with interactions that lead to the self-assembled structures that we observe in living
matter, including pseudo one-dimensional filaments, two-dimensional membranes,
tube-like channels and pores, and shell-like capsules.  Fascinating examples in the
latter category are the capsids of viruses, many of which are spheroidal and
are constructed from a specific number of copies of a small number of different
proteins \cite{Crick56a}.  It has long been known that the capsomers of some
viruses can assemble {\it in vitro}
into empty shells even in the absence of the viral genetic material \cite{Bancroft67a},
providing a powerful demonstration of how molecular shape and interactions dictate the
structure of aggregates.  For proteins, the symmetry and binding in oligomers is determined
by the contacts between neighbours in the complex, and it is now becoming possible
to engineer the quaternary structure by modifying the contact surface through mutations of
the amino acid sequence \cite{Grueninger08a}.
\par
Drawing inspiration from Nature, and encouraged by rapid advances in the synthesis of
tailor-made building blocks, computational scientists are trying to understand the
principles of self-assembly and how these can be used to build designed structures.  Explicit
models of polyhedral shell assembly have shown how difficult this objective can be \cite{Rapaport04a}.
While entropic and kinetic considerations are undoubtedly crucial for successful self-assembly,
there is also a clear requirement for target structures to be energetically stable
and kinetically accessible \cite{Wales05a,Wales06a}.
A logical starting point is therefore to design building blocks with attractive sites or
``patches'' in geometries compatible with the target structure.  Monodisperse
discrete objects, as well as continuous structures, such as sheets, can be constructed
in this way \cite{Zhang04a,Wilber07a}.
\par
Highly directional interactions can also be achieved through a non-uniform charge
distribution in particles that remain neutral overall.  For example, dipolar particles
will endeavour to form chains, since the head-to-tail arrangement of two dipoles is
low in energy and mechanically stable.  The tendency to form chains is partially
frustrated in the presence of competition for more compact arrangements arising
from an isotropic van der Waals or depletion attraction \cite{Clarke94a,VanWorkum05a}.  A simple
model incorporating these features is the Stockmayer potential, consisting of particles
with a Lennard-Jones (LJ) site plus a central point dipole.  By adjusting the
relative strength of the LJ and dipolar contributions, the energetically most stable
morphology of the 13-particle Stockmayer cluster changes in four stages from a distorted
icosahedron to a closed ring \cite{Clarke94a,Oppenheimer04a}.  For slightly larger sizes,
knots, links and coils emerge \cite{Miller05a}.
\par
Van Workum and Douglas have investigated the self-assembly of chains in low-density
Stockmayer fluids \cite{VanWorkum05a}, regarding the process as a form of reversible
polymerisation.  These authors have also considered a natural extension of the Stockmayer
potential to higher multipoles, in particular an LJ site plus point
quadrupole \cite{VanWorkum06a,Douglas07a}.  Quadrupole--quadrupole interactions favour
the formation of extended two-dimensional sheets, which can produce tubes when the edges
become connected.  The LJ-plus-multipole class of potentials therefore provides control
over the preference for compact (three-dimensional), sheet-like (two-dimensional) and
chain-based (one-dimensional) structure in self-assembly.  These tendencies compete with
each other.  In the present contribution we provide a systematic survey of the structure of
small clusters of quadrupolar spheres in an attempt to identify and understand the structural
motifs that emerge from the frustration between the isotropic and directional components of
the potential.

\section{Methods}

\subsection{Model potential}

The quadrupolar sphere is modelled as an isotropic Lennard-Jones
site with a point quadrupole of variable strength superimposed
\cite{OShea97a,VanWorkum06a}.
The pair potential, which we denote as LJQ, is of the form
\begin{multline*}
V_{ij}({\bf R}_{ij},\Omega_i,\Omega_j) =
4u\left[\left(\frac{\sigma}{R_{ij}}\right)^{12}-\left(\frac{\sigma}{R_{ij}}\right)^6\right]\\
+V_{\rm Q}({\bf R}_{ij},\Omega_i,\Omega_j),
\end{multline*}
where ${\bf R}_{ij}$ is the vector from particle $i$ to $j$, $R_{ij}$ is the magnitude of this
vector, and $\Omega_i$ represents the orientational degrees of freedom
of particle $i$.  The parameters $u$ and $\sigma$ are the Lennard-Jones dimer 
equilibrium well depth and separation, respectively, and will be used as the units of energy and length
henceforth.  $V_{\rm Q}$ is
the quadrupole--quadrupole interaction, which depends on the component(s) of the quadrupole
tensor involved.
\par
\begin{figure}
\centerline{\includegraphics[width=80mm]{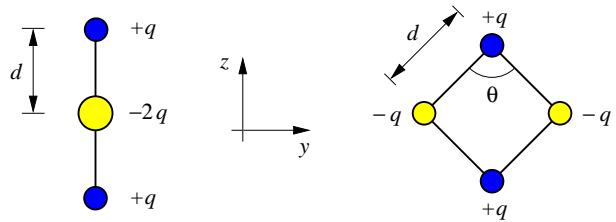}}
\caption{Definition of the charge distributions in the ``linear'' (left)
and ``square'' (right) quadrupoles with the local axis frame.
The point quadrupole is obtained in
each case by taking $d\to0$ while keeping $Q=qd^2$ fixed.
\label{quaddef}
}
\end{figure}

In this work, we consider the two quadrupolar arrangements of charges depicted in \fig{quaddef}.
In each case, the point quadrupole is reached by taking the limit in which the separation of the
charges $d$ goes to zero, while the strength $Q=qd^2$ of the quadrupole moment is
held fixed.
For the linear arrangement of charges, the interaction between two point quadrupoles $i$ and $j$
can be written in terms of the unit vectors ${\bf e}_{iz}$ and ${\bf e}_{jz}$ along the body-fixed
$z$-axes of the particles as
\begin{eqnarray}
V_{\rm Q}^{\rm lin}&=& 3(Q^*)^2u\left(\frac{\sigma}{R_{ij}}\right)^{5} \times \\
&& \big(1+2c_{zz}^2 - 20c_{zz}r_{iz}r_{jz}
- 5r_{iz}^2 - 5r_{jz}^2 + 35r_{iz}^2r_{jz}^2\big), \nonumber
\label{Vlin}
\end{eqnarray}
where $c_{\alpha\beta}={\bf e}_{i\alpha}\cdot{\bf e}_{j\beta}$, and
$r_{i\alpha}={\bf e}_{i\alpha}\cdot{\bf R}_{ij}/R_{ij}$,
$r_{j\alpha}={\bf e}_{j\alpha}\cdot{\bf R}_{ij}/R_{ij}$ (note the sign convention
with respect to the direction of ${\bf R}_{ij}$), with $\alpha$ and $\beta$
representing $x$, $y$ or $z$.
The dimensionless parameter $Q^*=Q/(4\pi\epsilon\sigma^5u)^{1/2}$
is the reduced quadrupole strength, $\epsilon$ being the dielectric permittivity
of the medium.
For the ``linear'' quadrupole defined by \eq{Vlin}, we represent the orientation
${\bf e}_{iz}$ of the quadrupole using spherical polar angles.
\par
In the $d\to0$ limit, the pair interaction for the square arrangement of charges in
\fig{quaddef} is
\begin{multline*}
V_{\rm Q}^{\rm squ}=\textstyle{\frac{3}{4}}(Q^*)^2u\left(\displaystyle\frac{\sigma}{R_{ij}}\right)^{5}
\big(2c_{yy}^2 - 2c_{yz}^2 - 2c_{zy}^2 + 2c_{zz}^2 \\
- 20c_{yy}r_{iy}r_{iy} + 20c_{yz}r_{iy}r_{jz} + 20c_{zy}r_{iz}r_{jy}\\ - 20c_{zz}r_{iz}r_{jz}
+35r_{iy}^2r_{jy}^2 - 35r_{iy}^2r_{jz}^2\\ - 35r_{iz}^2r_{jy}^2 + 35r_{iz}^2r_{jz}^2\big).
\end{multline*}
Three variables are now required to specify the orientation of the quadrupole, and
we have chosen to represent the vectors ${\bf e}_{iy}$ and ${\bf e}_{iz}$ in terms of
Euler angles.  The LJQ potential with $V_Q=V_{\rm Q}^{\rm squ}$ has received less
attention \cite{VanWorkum06a,Douglas07a} in the past than LJQ with $V_Q=V_{\rm Q}^{\rm lin}$.
\par
The linear quadrupole corresponds directly to the spherical tensor component $Q_{20}$.
The square quadrupole can be reached continuously from the linear arrangement by increasing
the angle $\theta$ in \fig{quaddef} from 0 to $\pi/2$, which corresponds to introducing
a contribution from the component $Q_{22c}$.  At $\theta=\pi/2$, the combination is
$\frac{3}{2}Q_{20}+\frac{1}{2}\sqrt{3}Q_{22c}$.  The interactions between components of the
quadrupole in the spherical tensor representation are tabulated in Appendix F
of Ref.~\cite{Stone97a}.
\par
We will need to perform local geometry optimisations using the LJQ potentials, and have
therefore derived and coded their analytic derivatives with respect to
the Cartesian position coordinates and the angular orientational variables.

\subsection{Global optimisation}

We performed unbiased searches for the global minima of clusters bound by the LJQ
potentials using the basin-hopping algorithm \cite{Wales97a}, in which a Monte Carlo
simulation is run on a transformed potential energy surface (PES)
by performing a local minimisation of the energy
at each step.  The local minimisation \cite{Li87a} is key to the success of 
basin-hopping \cite{Wales99a},
and is a feature shared by other efficient methods of global optimisation for clusters,
such as certain genetic algorithms \cite{Johnston03a}.
\par
For a given number of particles, $N$, and
quadrupole strength, $Q^*$, several runs seeded from different random initial positions
and orientations were performed.  The number of
Monte Carlo steps required to find a putative global minimum reliably in independent runs
depends strongly on the size of the cluster and the strength of the quadrupole moment.
It is also important to select a reasonable temperature for the accept/reject step
in the basin-hopping runs.  Although the
success of the method is not very sensitive to the temperature, it must be high
enough for the search to escape from local traps, but not so high that we fail to sample
the low-lying minima in each region sufficiently.
A fixed reduced temperature of $kT/u=1$ was often found to work well.
\par
To generate a structural map for the clusters, it is necessary to explore the two-dimensional
parameter space defined by the size of the cluster and the strength of the quadrupole.  The
straightforward approach of running basin-hopping on a grid of $Q^*$ points for each $N$
would be inefficient, since a small change in $Q^*$ will often lead only to a relaxation of the
global minimum, with qualitative changes to a new structure occurring at larger intervals
in $Q^*$.  We have therefore devised a surveying scheme with an iterative element, designed
to identify the values of $Q^*$ where the identity of the global minimum changes for a given $N$.
\par
The algorithm begins with thorough searches for the global potential energy
minimum at two values of the quadrupole strength,
$Q^*_{\rm low}$ and $Q^*_{\rm high}$,
that are far enough apart to lead to qualitatively different
structures.  These structures are then relaxed by local minimisation on a grid of $Q^*$ points
that lie between $Q^*_{\rm low}$ and $Q^*_{\rm high}$, resulting in the correlation of the energy
of each structure with $Q^*$.  During this process, it is possible that
one or both of the minima will disappear at some value of $Q^*$ due to a catastrophe in the
PES \cite{Wales01b}, leading to a sudden change in energy as the structure falls into a different
basin of attraction.  In this case, a new basin-hopping run is performed at the $Q^*$ where the
catastrophe occurred and the scan in $Q^*$ is continued.
\par
Eventually, the energies of the
relaxed structures initiated from $Q^*_{\rm low}$ and $Q^*_{\rm high}$ cross at
some quadrupole strength $Q^*_{\rm cross}$.  New basin-hopping runs are then performed
to identify the global minima at $Q^*_{\rm cross}\pm\delta Q^*$ a little above and below the
crossing point.  If the basin-hopping run at $Q^*_{\rm cross}-\delta Q^*$ returns the same
structure and energy as the relaxed structure from $Q^*_{\rm low}$, then we assume that this
structure was the global minimum not only at $Q^*_{\rm low}$ and at $Q^*_{\rm cross}-\delta Q^*$,
but also at all values of $Q^*$ in the intervening range.  In other words, we assume that there are no
reentrant global minimum structures.  This assumption is not only intuitively reasonable,
given that $Q^*$ continuously changes the potential from isotropic van
der Waals attraction to highly directional electrostatic interactions, but it is also borne
out by careful checks of particular cases.  The range $Q^*_{\rm cross}+\delta Q^*$ to
$Q^*_{\rm high}$ was treated analogously.  Since a local relaxation is much faster than a full basin-hopping
run, this procedure is far more efficient than using basin-hopping afresh at each intermediate $Q^*$
value.
\par
If, on the other hand, the basin-hopping runs at $Q^*_{\rm cross}\pm\delta Q^*$ return new
structures with lower energy
than the relaxed structures from $Q^*_{\rm low}$ and $Q^*_{\rm high}$, then this global minimum
supersedes them and was
in turn relaxed at values of $Q^*$ in both directions away from $Q^*_{\rm cross}$ until the
energy rose above those of the previous structures.  A new check for the true global minimum must
now be performed at this crossing point.  The procedure was terminated when no lower minima 
were found at the
crossing points of relaxed structures.  Hence, full basin-hopping runs need only be performed close
to the locations where the identity of the global minimum changes.

\section{Results}

\subsection{Local coordination of quadrupoles}

For a given separation ${\bf R}$, the energetically optimal arrangement of two point quadrupoles
is with the local $y$ axis on one particle and the local $z$ axis on the other aligned
with ${\bf R}$ and the other two local axes coplanar.  This is true for any value of $\theta$ in
\fig{quaddef}, but we will refer to the arrangement as a ``T-shape,'' which is most clearly
seen for the linear case, $\theta=0$.

\begin{figure}
\centerline{\includegraphics[width=70mm]{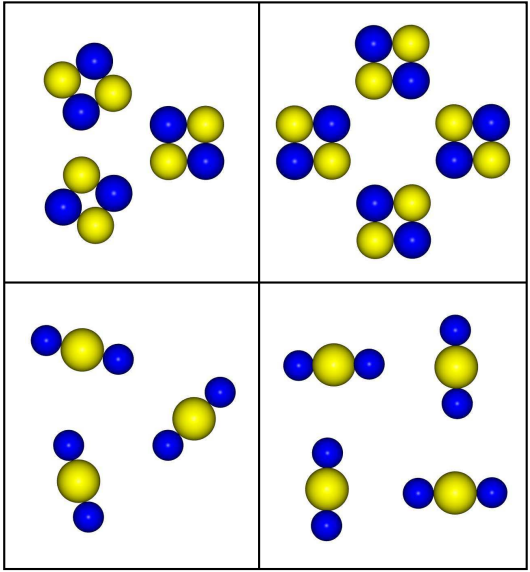}}
\caption{Characteristic coordination motifs for the linear (left) and square (right)
quadrupoles: global minima for the trimer (upper panels) and tetramer (lower) for a
strong quadrupole (large $Q^*$).  Extended charge arrangements are shown for
illustration only; all calculations are in the point quadrupole limit.
In the case of LQ$_4$ (lower left), the quadrupole axes are not coplanar.
\label{motifs}
}
\end{figure}

For the trimers, denoted LQ$_3$ and SQ$_3$ for the linear and square
quadrupoles, respectively, triangular arrangements are optimal, as shown
in the upper panels of \fig{motifs}.  Despite the significant distortion 
away from three ideal T-shaped pair interactions, the energies of the trimers
are each about $2.8$ times the respective dimer energies at $Q^*=5$.
\par
In the tetramers (lower panels of \fig{motifs}), the strain is relieved,
making four undistorted T-shapes possible.  For an interior angle of $45^\circ$,
the slipped-parallel arrangement of the next-nearest neighbours is also
favourable, further lowering the total energy of the tetramers to about $4.5$
times that of the respective dimers at $Q^*=5$.
An important difference between
the linear and square quadrupoles is demonstrated by the tetramers.  The axial
arrangement of charges in the linear quadrupole means that rotation of a quadrupole
about a local $z$ axis makes no difference to the energy.  LQ$_4$ is therefore
able to lower its energy by twisting the quadrupolar axes slightly.  This distortion
places diagonally opposite particles above and below the plane of the projection
in the lower-left panel of \fig{motifs}, allowing next-nearest neighbours to approach
more closely.  In contrast, SQ$_4$ does not have this flexibility, and the
structure in the lower-right panel of \fig{motifs} is completely planar.
\par
We note that the T-shape and slipped-parallel pair geometries are stationary
points for dimers of the linear quadrupolar molecule carbon dioxide.  However,
in contrast to the LJQ model, the T-shape of (CO$_2$)$_2$ is a saddle point,
while the slipped-parallel geometry is stable \cite{Bukowski99a}.

\subsection{Strong and weak quadrupole limits}

\begin{figure}
\centerline{\includegraphics[width=85mm]{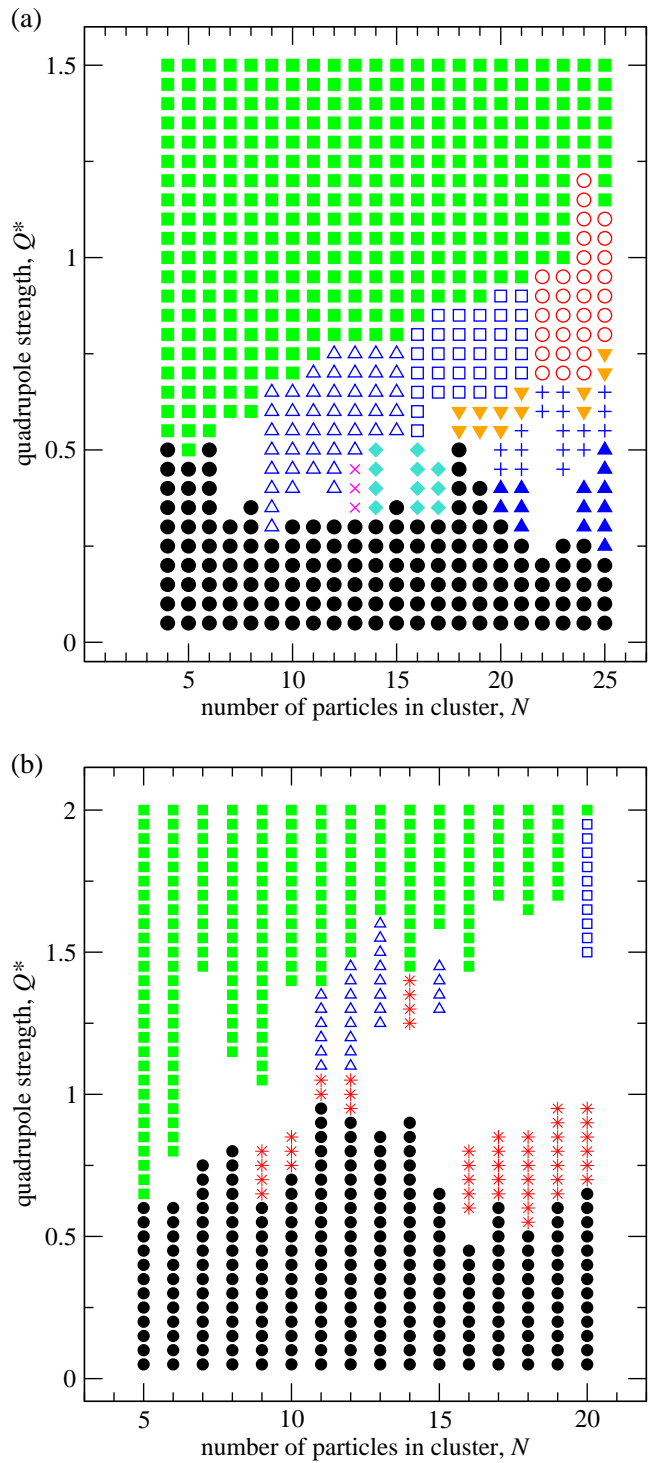}}
\caption{Structural maps for (a) the square and (b) the linear quadrupole:
relaxed LJ structure (filled circle), sheet (filled square),
stacked triangular antiprisms (open upward triangle), decorated triangular antiprisms
(filled upward triangle), stacked square antiprisms
(open square), filled stacked pentagonal antiprisms (plus), hollow shell (open circle),
filled shell (filled downward triangle), lattice-like (cross),
decahedral core (filled diamonds), star (star).
\label{maps}
}
\end{figure}

The T-shaped nearest-neighbour geometry favoured by the quadrupole--quadrupole
interactions encourages the formation of two-dimensional square networks.  However,
this tendency is frustrated by the LJ part of the potential, which drives the
structure towards compact, highly-coordinated arrangements with polytetrahedral
or icosahedral packing \cite{Wales97a}.  This competition produces a series
of structural motifs that partially satisfy the two opposing trends.  \fig{maps}
summarises the structural maps of SQ$_N$ and LQ$_N$ as a function of the number
$N$ of particles and the strength $Q^*$ of the quadrupole moment.  Some structures
are difficult to classify in an unambiguous or meaningful way, and such
combinations of $N$ and $Q^*$ have been left blank in the figure for clarity.

\begin{figure}
\centerline{\includegraphics[width=85mm]{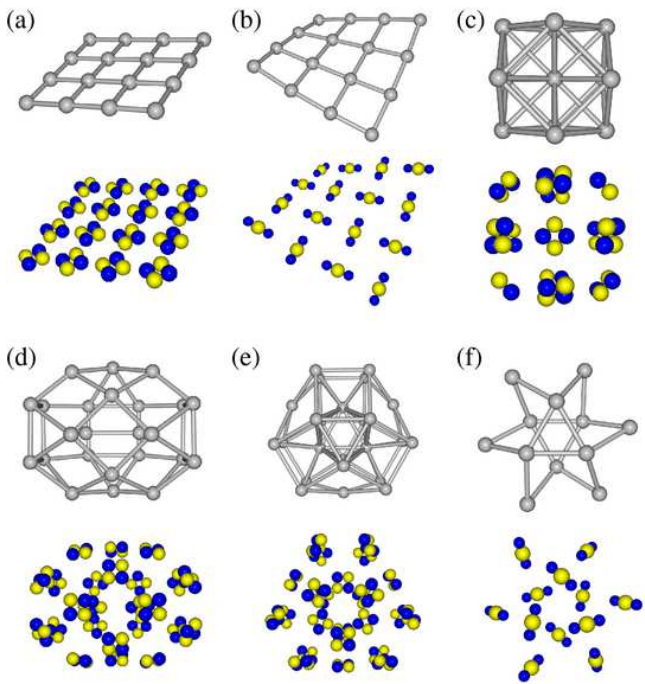}}
\caption{Structures discussed in the text.
(a) SQ$_{16}$ sheet at $Q^*=2$,
(b) LQ$_{16}$ sheet at $Q^*=1.5$,
(c) SQ$_{13}$ cuboctahedron at $Q^*=0.4$,
(d) SQ$_{24}$ filled shell at $Q^*=0.6$,
(e) SQ$_{24}$ decorated triangular antiprisms at $Q^*=0.4$,
(f) LQ$_{12}$ star at $Q^*=1.025$.
\label{structures}
}
\end{figure}

For a sufficiently weak quadrupole moment, the global minimum must be close to the
LJ global minimum structure, but with slight distortions induced by the quadrupoles.
However, effectively confining the quadrupoles to an icosahedral framework
for SQ$_{13}$
causes the quadrupoles to experience severe frustration, akin to that found in
geometrically frustrated magnets \cite{Harrison04a}.  Hence, a given arrangement of
the particles can correspond to multiple potential energy minima in the orientational
part of configuration space, introducing a new source of complexity to the PES.
The number of such isomers of the LJ structure generally increases with $Q^*$,
but also depends sensitively and non-monotonically on $N$.
For the near-icosahedral SQ$_{13}$ there are two distinct
isomers at $Q^*=0.025$, while for SQ$_{19}$, where the
global minimum is based on two interpenetrating icosahedra for
this value of $Q^*$, we located
23 distinct arrangements by quenches of the LJ structure starting from random quadrupole
orientations.  This figure rises to several hundred for SQ$_{19}$ at $Q^*=0.1$.
Although these searches are not definitively exhaustive, the rapid increase in
the number of isomers illustrates the roughness of the PES with respect to the
orientational coordinates.
\par
In the opposite
limit of large $Q^*$, a two-dimensional sheet always emerges.  The sheets for the
square number $N=16$ are shown in \fig{structures}.  The
twist seen in LQ$_4$ is continued as the sheet grows, while the sheets of square quadrupoles are
always planar.  The sheet grows by adding particles at adjacent sites of the extended square
lattice, with the dimensions of the lattice adapting to maximise the number of T-shaped
pairs in the first instance.  It is often possible to achieve the maximum number of such
pairs in more than one way on a square lattice, and next-nearest neighbour interactions
then come into play.  Hence, there can be close competition between structures even in
the strong quadrupole limit.  In contrast, the Stockmayer potential always has an unambiguous
optimal structure consisting of a planar ring of head-to-tail dipoles in the strong dipole
limit \cite{Miller05a}.
The lower boundary of the quadrupole sheet on the structural map
moves (non-monotonically) to higher $Q^*$ as $N$ increases,
because a larger number of LJ pair interactions must be disrupted to create the sheet.
The widening region between the relaxed LJ cluster and the sheet is occupied
by structures that strike a compromise between high-coordination number and sheet-like
arrangements.

\subsection{The 13-particle cluster}

\begin{figure}
\centerline{\includegraphics[width=85mm]{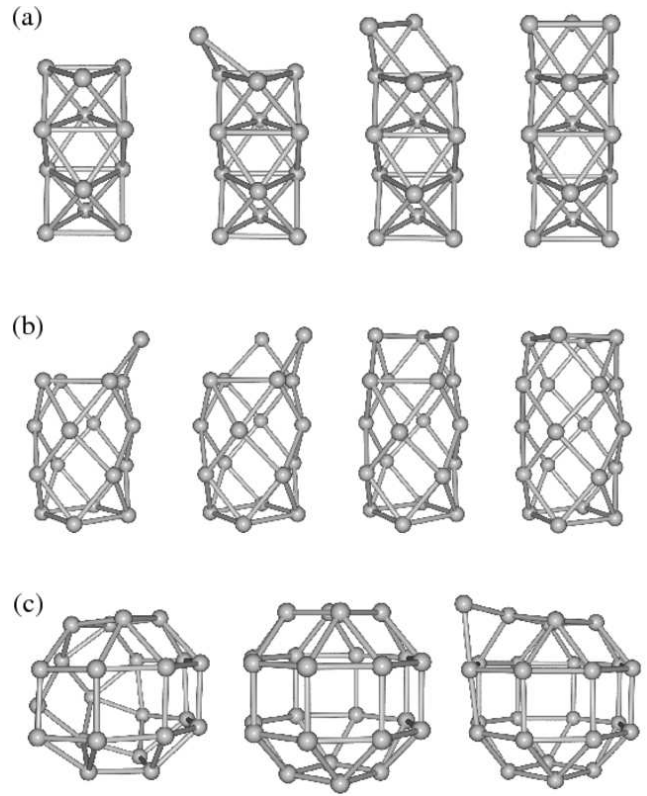}}
\caption{Part of some growth sequences for the square quadrupole.  (a) stacked triangular antiprisms,
(b) stacked square antiprisms, (c) hollow shells.
\label{growth}
}
\end{figure}

The 13-particle cluster, which for the pure LJ potential has a global minimum
consisting of a centred icosahedron with point group $I_h$, provides a good
illustration of the sequence of changes driven by the quadrupole in small clusters.
For SQ$_{13}$ with small $Q^*$, the quadrupole of the central particle
aligns itself perpendicular to one of the icosahedral $C_2$ axes.  The resulting small
distortions of the particle positions lower the symmetry to $C_{2h}$.
The quadrupole--quadrupole interactions are highly frustrated when
confined to the vertices of the distorted icosahedron and at $Q^*=0.35$, the global
minimum switches to a slightly distorted centred cuboctahedron ($D_{2d}$)---a fragment of
face-centred cubic lattice (\fig{structures}c).  This
structure maintains the 12-fold coordination of the central
particle but the shell consists of squares and triangles, with quadrupole arrangements
closer to those in the right-hand panels of \fig{motifs}.  At $Q^*=0.525$, a second change occurs,
to a stack of face-sharing triangular antiprisms (illustrated in the second panel of
\fig{growth}a), which can also be regarded as a narrow tube if viewed down the three-fold
axis.  This structure sacrifices the high coordination of a more spherical lattice-like
fragment for an elongated arrangement, in which the quadrupoles are better aligned.
The switch to the sheet structure then takes place in two steps.  First, at $Q^*=0.8$ a
$3\times4$ sheet arises with the thirteenth particle in the same plane, bridging the
central bond of a long side.  At $Q^*=1.15$, the triangular face becomes too unfavourable,
and the sheet adopts a $4\times4$ square with three of the corners missing.

\begin{figure}
\centerline{\includegraphics[width=80mm]{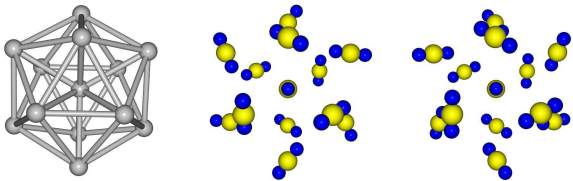}}
\caption{The 13-particle icosahedron (point group $I_h$) viewed along a $C_3$
axis and the two LQ$_{13}$ isomers at $Q^*=0.025$ of lowest energy.
Both belong to point group $S_6$.
\label{orientations}
}
\end{figure}

A related sequence emerges with $Q^*$ for the cluster of 13 linear quadrupoles.  Like the
dipole moment in the 13-particle Stockmayer cluster, the axis of the quadrupole on the central
particle in LQ$_{13}$ selects one of the $C_3$ axes of the icosahedron,
here reducing the symmetry from $I_h$ to $S_6$.  However, there are four
other icosahedral minima differing by the quadrupole orientations.  The two isomers
with the lowest energy are illustrated in \fig{orientations}.
With increasing $Q^*$, the cluster eventually passes to the stacked triangular
antiprisms, but does so through a rather amorphous structure, unlike the highly symmetric
cuboctahedron seen in SQ$_{13}$.  The linear quadrupole can tolerate a considerably higher
$Q^*$ before switching to the sheet than can the square quadrupole---an observation that
holds for all $N$ studied here.

\subsection{Structural families and growth sequences}

A number of structural families emerge in LJQ clusters and persist over some range of $Q^*$ and
$N$.  The stacked triangular antiprisms described above for 13 particles are seen for SQ$_N$
with $9\le N\le15$.  When $N$ is not a multiple of three, the end particles form an incomplete layer,
giving rise to a simple growth sequence, part of which is depicted in \fig{growth}a.  This
family does not continue indefinitely as the global minimum,
but is replaced for $16\le N\le21$ by stacked square
antiprisms (\fig{growth}b).  The diamond-like faces on the surface of the square-based structure
are flatter and closer to the ideal tetramer arrangement than those on the surface of the
triangular antiprismatic stack.
\par
Increasing the circumference of these stacks by another particle to make pentagons makes
the diameter of the structure large enough to accommodate a line of particles down the centre
of the stack, giving a filled tube-like arrangement.
The additional contacts provided by the central line make such structures competitive at
lower $Q^*$ than the stacked squares.
In fact, pentagonal stacks arise in two forms in the structural map.  The diamond symbols in
\fig{maps}(a) indicate clusters built around a decahedron, which contains
a pentagonal prism.  For pair potentials with a preferred nearest-neighbour separation,
decahedral structures are less strained than icosahedral ones and are
seen in the global minima of short-ranged isotropic potentials \cite{Doye95d} and
experimentally in metal clusters \cite{Marks94a}.
In LJQ clusters, the square faces are favoured by the quadrupolar interactions, and
the structure can grow by building additional partial layers around the five-fold axis.
Twisting the pentagonal layers gives the second type of pentagonal stack structure,
filled pentagonal antiprisms, which appear on the map for SQ$_N$ with $N\ge20$.  
These tube-like stacks can be regarded as two-dimensional
sheets in which two opposite edges have been joined, thereby exchanging the energetic cost of an exposed
edge for the penalty of curving the sheet.  This trade-off is analogous to the formation of
closed rings of dipoles \cite{Miller05a}.  Larger tubes have been observed to assemble spontaneously
in the finite-temperature simulations of the LJQ fluid by Van Workum and Douglas \cite{VanWorkum06a}.
\par
Tubes can dispose of their remaining exposed edges by also closing the ends to make a
shell.  We observe hollow shells over a range of $Q^*$ in SQ$_{22}$ and larger.
For this class of structures, certain values of $N$ give rise to a structure of high symmetry.
The first of these is SQ$_{24}$, illustrated in the central panel of \fig{growth}c, which
has perfect $O_h$ octahedral symmetry.  The shell normally grows by insertion of a particle into
the surface, causing a distortion of the ideal triangular and square faces, but occasionally by
the addition of an edge-bridging particle, as shown in \fig{growth}c.  The shell seems to be a
permanent feature of larger SQ$_N$ clusters.  We have followed it as far as $N=36$, which
forms an elongated shell of $D_{3d}$ symmetry with triangular faces at the ends and
antiprismatically stacked hexagons along the body.
\par
If $Q^*$ is not sufficiently large, the shell is energetically penalised for its
shortage of LJ nearest neighbour pairs.  However, a large number of such pairs can be
obtained by placing a few particles inside the shell.  Hence, the hollow shell is
typically preceded by a filled shell or filled pentagonal tube in the structural
map, \fig{maps}a.  A shell
encapsulating two particles is shown in \fig{structures}d for SQ$_{24}$; compare the
hollow shell for this cluster in the central panel of \fig{growth}c.
\par
Similar energetic compromises produce mixtures of structures that have already been
described.  For example, the transition
from the LJ structure to the filled tubes and shells is sometimes bridged by decorated
versions of the stacked triangular antiprisms, where the stack has been surrounded by a
new layer.  This arrangement is illustrated in a view down the three-fold axis
for SQ$_{24}$ in \fig{structures}e.  The characteristic network of square faces
for the shell is beginning to emerge on the outside of these structures, but they maintain
a larger number of LJ pairs than the filled shell.
\par
The linear quadrupole tends not to give rise to hollow global minima.  Although tube-like and
shell-like structures do appear, they are collapsed into what would be the central space in the
square quadrupole equivalents.  This ability to distort, or inability to support a hollow
interior, is a result of the axial symmetry of the linear quadrupole.  A square array
of linear quadrupoles,
such as the one depicted in \fig{structures}b, can fold and twist along one of its
diagonals without severely disrupting the T-shaped nearest-neighbour interactions either
side of the fold.  This is not true of an array of square quadrupoles, such as that in
\fig{structures}a, in which each quadrupole defines a plane and not just a line.
By collapsing inwards, the clusters of linear quadrupoles gain
favourable interactions between opposite sides of the structure that would otherwise be
held far apart.  However, the collapsed structures are often rather amorphous, making them
hard to classify or describe in a helpful way.  For this reason, the structural map
in \fig{maps}b extends only to LQ$_{20}$.
\par
A distinctive and reproducible feature of the linear quadrupole clusters is a family
of structures with a star-like organisation of
the particles and a gear-wheel array of quadrupole axes, which are scattered around the structural
map (\fig{maps}b).  Again, for particular values of $N$, the cluster can achieve a high symmetry
that may be based on a three-fold or four-fold principal symmetry axis.  An example belonging
to point group $S_6$ is shown in \fig{structures}f.  However, the stability of these morphologies
is strongly correlated with the number of particles; the stars are not
observed away from the values of $N$ that allow the symmetry to be completed.

\section{Concluding Remarks}

The survey of putative global minima presented here shows that the competition between
isotropic attractive forces and highly directional quadrupole--quadrupole interactions
gives rise to a wide variety of structural motifs.  These include elongated tube-like
structures, hollow and filled shells, stars, and extended sheets.
Some unusual point groups are represented in this collection.
\par
Global optimisation is most challenging when the competing influences are closely balanced,
i.e., for intermediate strengths of the quadrupole in this work.  Obtaining reproducible
lowest-energy structures was significantly more difficult for the quadrupolar potential than for
equivalent dipolar Stockmayer potential \cite{Miller05a}.
This observation, together with the multiplicity of minima
that were found to have virtually identical positions but different orientations of the quadrupoles,
hints at a complex potential energy surface
in certain parts of the $(N,Q^*)$ parameter space.  Confirmation
and further exploration of this complexity would require a more comprehensive analysis of the
energy landscape \cite{Wales00b}.  The landscape approach would provide information not only on the
number of competing structures, but also on the barriers separating them and the rearrangement
mechanisms that interconvert them.  This information would provide a starting point for investigating the
thermal stability and dynamic properties of the clusters, as well as the routes by which they
might self-assemble.
\par
The various families of structures have their own ``magic'' numbers at which a particular
shape is complete and the landscape is probably minimally frustrated \cite{Bryngelson87a,Wales06a}.
Such numbers are well known for a variety of simpler interatomic potentials,
and often correspond to the completion of successive
icosahedral shells \cite{Mackay62a} at $N=13,\ 55,\ 147\dots$.  The special stability associated
with these sizes, combined with kinetic accessibility \cite{Wales06a},
can lead to prominent features such as experimental abundance \cite{Farges88a}.
In the present work, the tube-like structures of
stacked triangular and square antiprisms achieve completed layers for multiples of three and
four particles, respectively, while a shell can be elongated by the insertion of a complete
hexagonal ring.  In the strong quadrupole regime, sheets adopt defect-free squares when $N$
is a square number.  It would be interesting to investigate whether these ``perfect'' structures
are especially stable and self-assemble efficiently, as for magic number Lennard-Jones
clusters \cite{Doye99c}.
\par
We have seen that quadrupole--quadrupole interactions favour the formation of extended two-dimensional
structures with four-fold coordination of the particles.  In contrast, dipole--dipole interactions
lead to extended pseudo one-dimensional chains, while isotropic attraction drives the structure towards
compact three-dimensional arrangements.  By careful balancing of the multipolar interactions, it
should therefore be possible to exert considerable control over the structures that self-assemble out
of multipolar particles with isotropic core interactions.
\par
Briefly considering bulk phases, rather than finite clusters,
such control could be useful in adjusting the networking properties of colloidal gels.  For example,
it has recently been shown that dipolar colloids can be encouraged to form more interconnected
gel-like networks by a slight extension of the dipole \cite{Blaak07a}.  From studies
of models of patchy spheres with fixed maximum valency it is now known that the average coordination
number of the particles in a gel has important consequences for the structure of the gel and for the
underlying phase behaviour of the fluid from which it forms \cite{Bianchi06a}.  The present
work suggests that the average coordination number could be finely
tuned either by adding a weak point quadrupole
to point dipolar particles, or by using a mixture of dipolar and quadrupolar spheres.
Hence, multipolar particles could be an appealing alternative to patchy colloids for realizing
and exploring reversible gels \cite{Zaccarelli07a}.

\acknowledgments

The authors are grateful to Josef O'Brien for some preliminary calculations on Lennard-Jones
clusters with extended quadrupolar distributions of point charges.  MAM thanks EPSRC for
financial support.


\end{document}